\documentclass[twocolumn,floatfix,showpacs]{revtex4}%
\usepackage{graphicx}%
\usepackage{amsmath}%
\usepackage{amsfonts}%
\usepackage{amssymb}
\usepackage{bm}

\def\s{{\sigma}}
\def\e{{\epsilon}}
\def\k{{ {\bm k} }}

\def\q{{ {\bm q} }}

\def\w{{\omega}}
\def\a{{\alpha}}

\allowdisplaybreaks[0]

\begin{document}
\title{
Revisit of the Orbital-Fluctuation-Mediated Superconductivity in LiFeAs: \\
Nontrivial Spin-Orbit Interaction Effects on the Bandstructure \\
and Superconducting Gap Function
}
\author{
Tetsuro \textsc{Saito}$^{1}$, 
Youichi \textsc{Yamakawa}$^{1}$, 
Seiichiro \textsc{Onari}$^{2}$, and
Hiroshi \textsc{Kontani}$^{1}$,}

\date{\today }

\begin{abstract}

Precise gap structure in LiFeAs ($T_{\rm c}=18$K)
given by ARPES studies offers us significant information
to understand the pairing mechanism in iron-based superconductors.
The most remarkable characteristics in LiFeAs gap structure would be that 
``the largest gap emerges on the tiny hole-pockets around Z point''.
This result had been naturally explained in terms of the 
orbital-fluctuation scenario 
(T. Saito {\it et al}., Phys. Rev. B {\bf 90}, 035104 (2014)),
whereas an opposite result is obtained by the spin-fluctuation scenario.
In this paper, we study the gap structure in LiFeAs
by taking the spin-orbit interaction (SOI) into account,
motivated by the recent ARPES studies that revealed
the significant SOI-induced modification of the Fermi surface topology.
For this purpose, we construct the two possible tight-binding models 
with finite SOI by referring the bandstructures given by different ARPES groups. 
In addition, we extend the gap equation for multiorbital systems with finite SOI, 
and calculate the gap functions by applying the orbital-spin fluctuation theory. 
On the basis of both SOI-induced band structures, 
main characteristics of the gap structure in LiFeAs are naturally reproduced 
only in the presence of strong inter-orbital interactions 
between ($d_{xz/yz}$-$d_{xy}$) orbitals.
Thus, the experimental gap structure in LiFeAs is a strong 
evidence for the orbital-fluctuation pairing mechanism.

\end{abstract}

\address{
$^1$ Department of Physics, Nagoya University,
Furo-cho, Nagoya 464-8602, Japan. 
\\
$^2$ Department of Physics, Okayama University,
Okayama 700-8530, Japan.
}
 

\sloppy

\maketitle

\section{Introduction}
The pairing mechanism of iron-based superconductors 
had been studied intensively as a central issue.
Up to now, the spin-fluctuation-mediated $s_{\pm}$-wave state with sign reversal
\cite{Kuroki,Chubukov,Mazin,Hirschfeld,Thomale} and
orbital-fluctuation-mediated $s_{++}$-wave state without sign reversal
\cite{Kontani-RPA,Onari-SCVC,Review} had been proposed.
According to theoretical studies for realistic multiorbital models,
$s_\pm$-wave state is fragile against impurities:
$s_\pm$-wave state with $T_{\rm c}=30$K disappears
when the residual resistivity $\rho_0$ is as large as 
$\sim15 \ \mu\Omega{\rm cm}$ for $m^*/m_b\sim3$ 
by $3d$-transition metal impurities \cite{Onari-imp,Yamakawa-imp}.
Experimentally, the superconductivity in LaFeAsO+F,
(K,Ba)Fe$_2$As$_2$ and Ba(Fe,Co)$_2$As$_2$ survives till
$\rho_0 \sim 300 \ \mu\Omega{\rm cm}$  \cite{Sato-imp,Li-imp,Nakajima-imp,Taen},
indicate that the $s_{++}$-wave state occurs in these compounds.
On the other hand, $T_c=18$K disappears only 
when $\rho_0\sim 40 \ \mu\Omega{\rm cm}$ in Ba(Fe,Ru)$_2$As$_2$
by electron irradiation measurement \cite{Prozorov}.
In the neutron spectroscopy, the resonance-like peaks
in various compounds
are also explained in terms of the $s_{++}$-wave state
by taking the realistic quasiparticle inelastic scattering 
into account \cite{Onari-neutron}.

In many Fe-based superconductors, the gap structure has been determined
in detail by the angle-resolved photoemission spectroscopy (ARPES) studies.
Although ARPES is not a sign-sensitive experiment,
the detailed gap structure given by the ARPES
offers us very useful information to distinguish the pairing mechanism.
In fact, orbital dependence of the gap function is very crucial
since it reflects the orbital dependence of the pairing interaction strength.
In the spin-fluctuation mechanism,
the superconductivity is mainly induced by the intra-orbital scattering 
of the Cooper-pairs, since the spin-fluctuations are driven by the 
intra-orbital Coulomb interaction $U$.
For this reason, the gap magnitude $|\Delta(\k)|$ tends to largely depend on the 
orbital character of the Fermi surface in the spin-fluctuation mechanism.
In contrast, in the orbital-fluctuation mechanism, 
the superconductivity is induced by the inter-orbital scattering 
of the Cooper-pairs.
For this reason, the orbital dependence of $|\Delta(\k)|$ 
tends to be moderate in the orbital-fluctuation mechanism.

Motivated by the above discussion,
the gap magnitude on the  $d_{z^2}$-orbital hole-pocket around Z point
in Ba122 and Sr122 compounds had been studied intensively.
Because of the absence of the intra-$d_{z^2}$-orbital nesting,
the spin-fluctuation scenario predicts the emergence of the 
''horizontal node'' on the hole-pocket around Z point \cite{Suzuki,Saito-loop}.
In contrast, the horizontal node is absent in the orbital-fluctuation-mediated 
$s$-wave state, since the $d_{z^2}$-orbital electrons contribute to 
the orbital-fluctuations 
\cite{Saito-loop}.
In BaFe$_2$(As,P)$_2$ and Ba(Fe,Co)$_2$As$_2$,
several ARPES groups reported the absence of the horizontal node
\cite{Shimo-Science,Yoshida,Hajiri},
whereas horizontal node was reported in Ref. \cite{Zheng}.

\begin{figure}[!htb]
\includegraphics[width=0.99\linewidth]{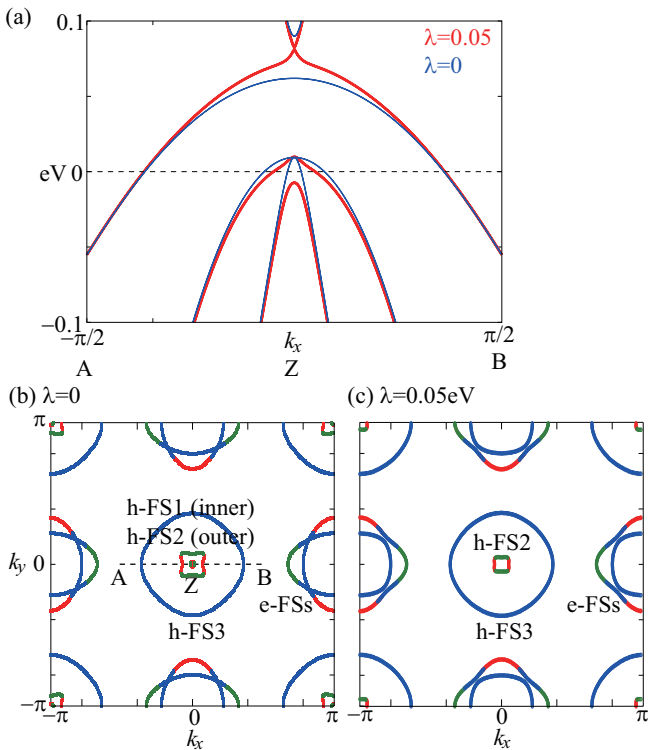}
\caption{
(Color online)
(a) The dispersion of the band structure between points A and B 
for $\lambda = 0.05$ eV (red lines) and $0$ (blue lines), respectively.
We call the former the ``SOI-induced bandstructure (I)''.
A and B points are shown in (b).
The corresponding FSs for (b) $\lambda = 0$ and (c) $0.05$ eV 
in the $k_z = \pi$ plane are shown respectively.
In (c), h-FS1 disappears due to the SOI.
Green, red, and blue lines correspond to $xz$, $yz$, and $xy$ orbitals, 
respectively.
}
\label{fig:FS}
\end{figure}

For studying the pairing mechanism,
LiFeAs is an suitable compound since the detailed band structure and
the superconducting gap functions had been determined by 
several ARPES groups \cite{Borisenko-LiFeAs,Ding-LiFeAs}.
LiFeAs is an ideal system for ARPES studies in that 
the charge neutral cleavage plane exists and 
clean single crystals can be synthesized.
Therefore, the intrinsic gap structure free from extrinsic effects
(such as the impurity effect) is obtained by ARPES studies.
The experimental band structure and Fermi surfaces (FSs) 
are shown in Fig. \ref{fig:FS}:
Reflecting the bad nesting between the hole-like FSs (h-FSs) 
and electron-like FSs (e-FSs) in LiFeAs, 
moderate spin-fluctuations are observed by
NMR studies \cite{NMR-LiFeAs}
and neutron scattering studies
\cite{neutron-LiFeAs1,neutron-LiFeAs2,neutron-LiFeAs3}.
Interestingly, in LiFeAs, the largest gap appears in 
the tiny three-dimensional (3D) hole-pockets (h-FS1,2) 
according to the ARPES studies
\cite{Borisenko-LiFeAs,Ding-LiFeAs}
and the Scanning Tunneling Microscopy (STM) study \cite{QPI}.

Based on the tight-binding model for LiFeAs constructed from the ARPES 
band structure, some authors studied the possible pairing scenarios.
In the spin-fluctuation mechanism for the $s_{\pm}$-wave state,
the gaps on the tiny hole-pockets,
h-FS1,2 in Fig. \ref{fig:FS} (b), are very small,
because of the bad intra-orbital nesting between h-FS1,2 and e-FSs
\cite{Hirschfeld-LiFeAs,Saito-LiFeAs}.
In contrast, 
in the orbital-fluctuation mechanism for the $s_{++}$-wave state,
the gaps on the tiny hole-pockets are the largest,
because of the good inter-orbital nesting 
\cite{Saito-LiFeAs}.
The latter theoretical result is consistent with experiments
\cite{Borisenko-LiFeAs,Ding-LiFeAs,QPI}.

However, although the SOI had been neglected
in previous theoretical studies in Refs.
 \cite{Hirschfeld-LiFeAs,Saito-LiFeAs},
recent ARPES studies for LiFeAs revealed that the SOI is never negligible
to construct the FSs and the band structure near the Fermi level.
For instance, the degeneracy of the $d_{xz/yz}$ hole bands 
on the $\Gamma$-Z line is lifted by the SOI \cite{Borisenko-SOI,Ding-SOI,Ding-Rev}.
For this reason, one of the tiny two hole-pockets (h-FS1,2)
disappears due to the SOI, as shown in  Fig. \ref{fig:FS} (c)
(or Fig. \ref{fig:newband-Li} (d)).
Similar disappearance of h-FS1 due to SOI
had been observed in FeSe \cite{FeSe1,FeSe2,FeSe3,FeSe4,FeSe5,FeSe6}.
Since the tiny hole-pockets play important roles in the pairing 
mechanism, a theoretical study for LiFeAs including the SOI 
had been highly required.

In this paper, we study the gap structure of LiFeAs
by taking the effect of the SOI into account,
in order to distinguish the correct pairing scenario.
For this purpose, we construct two possible 3D
ten-orbital models,
which correspond to the different experimental band structures
reported in Refs. \cite{Borisenko-SOI,Ding-SOI}.
Both models possess a single tiny hole-pocket around Z point.
We also introduce both the multiorbital Coulomb interaction ($U,U',J$)
and the quadrupole interaction ($g$):
The latter will originate from the 
Aslamazov-Larkin type vertex correction (AL-VC),
which describes the many-body effect beyond the mean-field approximation
\cite{Onari-SCVC,Review,Yamakawa-Raman,Tsuchiizu}.
By using the random-phase-approximation (RPA), 
the orbital- and spin-fluctuations develop with increasing $g$ and $U$, 
respectively.
Then, we extend the gap equation for multiorbital systems with finite SOI,
and calculate the gap function
by using the pairing interaction given by the RPA.
In the $s_{++}$-wave state due to orbital-fluctuations,
the gap on the smallest h-FS is the largest of all.
In contrast, in the $s_\pm$-wave state due to spin-fluctuations, 
the gap on the smallest h-FS is almost zero.
Therefore, the superconductivity in LiFeAs originates from the 
inter-orbital ($d_{xz/yz}$-$d_{xy}$) interactions due to orbital-fluctuations,
not from the intra-orbital spin-fluctuation interactions.

When both the orbital- and spin-fluctuations develop comparably,
we frequently obtain the ``hole-$s_{\pm}$-state'', in which the 
gap structure has the ``sign reversal between h-FSs''.
This interesting gap structure had been discussed in 
(Ba,K)Fe$_2$As$_2$ experimentally \cite{Watanabe-shpm,Zhang-shpm}.
In Appendix A, we show that the hole-$s_{\pm}$-wave gap structure 
can be obtained in the 3D model for Ba122 system.

\section{Formalism}
\label{sec:Formalism}

\subsection{SOI-induced bandstructure (I) and interaction terms}
In this paper, we set $x$ and $y$ axes parallel to the nearest Fe-Fe bonds,
and the orbital $z^2$, $xz$, $yz$, $xy$, and $x^2-y^2$ 
of Fe-A (B) site are denoted as
1, 2, 3, 4, and 5 (6, 7, 8, 9, and 10) respectively.
We used the 3D ten-orbital tight-binding model
which is given in Ref. \cite{Hirschfeld-LiFeAs}.
This model is obtained by fitting the experimentally observed dispersion
reported in Ref. \cite{Borisenko-LiFeAs}.
Note that the band renormalization due to 
the mass enhancement $m^*/m_b=2\sim3$ is taken into account
in the present model.

We stress that the experimental FSs of LiFeAs in Fig. \ref{fig:FS},
on which the present study is based,
are very different from the FSs given by the density functional theory (DFT), 
in which h-FS1,2 predicted by the DFT are much larger.
Better agreement between theory and ARPES is achieved 
by the LDA+DMFT study \cite{DMFT}, since the FS1,2
shrinks due to the orbital-dependent self-energy that is absent in the LDA.

In the absence of the SOI, 
the kinetic term of the ten-orbital model is given as
\begin{eqnarray}
\hat{H}^{0}&=& \sum_{a b l m \sigma} t_{l,m}(\bm{R}_{a}-\bm{R}_{b}) 
c^{\dagger}_{l \s}(\bm{R}_{a}) c_{m \s} (\bm{R}_{b})
 \nonumber \\
&=& \sum_{\k l m \s} \left\{ \sum_{a} t_{l,m}(\tilde{\bm{R}}_{a}) 
e^{i \bm{k} \cdot \tilde{\bm{R}}_{a}} \right\} c^{\dagger}_{l \s}(\bm{k}) c_{m \s} (\bm{k})
\label{eqn:H0}
\end{eqnarray}
where $l,m = 1-5$ ($6-10$) represent the $d$ orbitals at Fe-A (Fe-B) atom, and
$\s = \pm 1$ is the spin index.
$\tilde{\bm{R}}_{a}$ is the Fe-site position measured from an Fe-A atom,
$c^{\dagger}_{l \s}(\bm{k})$ is the creation operator of the $d$ electron,
and $t_{l,m}(\tilde{\bm{R}})$ is the hopping integral.
The values of $t_{l,m} (\tilde{\bm{R}})$ are shown in Ref. \cite{Hirschfeld-LiFeAs}.

Next, we introduce the interaction terms.
We consider both the multiorbital Coulomb interaction 
($U$, $U'$, $J=(U-U')/2$) and the quadrupole interaction.
The quadrupole interaction Hamiltonian is given as
\begin{eqnarray}
V_{\rm{quad}}=- g(\w_l) \sum_a^{\rm site} \left( 
{\hat O}^a_{yz} \cdot{\hat O}^a_{yz} + {\hat O}^a_{xz} 
\cdot{\hat O}^a_{xz} \right) 
 \label{eqn:Hint}
\end{eqnarray}
where $g(\w_l)=g \w_0^2/(\w_l^2+\w_0^2)$, and 
$g=g(0)$ is the quadrupole interaction at $\omega_l=0$.
$\omega_0$ is the cutoff energy of the quadrupole interaction.
$\hat{O}_{\Gamma}^a=\sum_{l,m}o^{l,m}_\Gamma {\hat m}_{l,m}^a$ 
(${\hat m}_{l,m}^a= \sum_\s c_{l\s}^\dagger(\bm{R}_a) c_{m\s}(\bm{R}_a)$)
is the quadrupole operator at site $\bm{R}_a$ introduced 
in Ref. \cite{Kontani-RPA}:
The non-zero coefficients of $o^{l,m}_\Gamma=o^{m,l}_\Gamma$ are 
$o_{xz}^{2,5}=o_{xz}^{3,4}=\sqrt{3}o_{xz}^{1,2}=1$, and
$-o_{yz}^{3,5}=o_{yz}^{2,4}=\sqrt{3}o_{yz}^{1,3}=1$.
Thus, $\hat{V}_{\mathrm{quad}}$ has many non-zero inter-orbital elements.
As explained in Ref. \cite{Kontani-RPA},
$g$ is given by in-plane Fe-ion oscillations.
Also, the AL-VC due to Coulomb interaction gives large effective 
quadrupole interaction $g$ \cite{Onari-SCVC}.

Now, we analyze the present model by applying the RPA.
The $32 \times 32 \times 16 \ \bm{k}$ meshes are used in the numerical study.
We fix the temperature at $T=0.01$ eV, 
the cutoff energy at $\omega_0 = 0.02$ eV,
and set the filling of each Fe-site as $n=6.0$.
The irreducible susceptibility is given as
{%
\begin{eqnarray} 
\chi^0_{ll',mm'} \left( q \right) =
- \frac{T}{N} \sum_k G_{l,m} \left( k+ q \right) G_{m',l'} \left( k \right),
\end{eqnarray}
where $q = ( \bm{q}, \omega_l )$ and $k=( \bm{k} , \epsilon_n)$.
$\epsilon_n = (2n + 1) \pi T$ and $\w_l = 2l \pi T$ 
are the fermion and boson Matsubara frequencies.
$\hat{G} ( k ) = [ i \epsilon_n + \mu - \hat{h}^0(\bm{k}) ]^{-1}$
is the Green function in the orbital representation,
where $\hat{h}^0(\bm{k})$ is the matrix representation of $\hat{H}^0$
for the momentum $\bm{k}$, and $\mu$ is the chemical potential.
In the RPA, the spin and charge susceptibilities in the matrix form
are given by 
\cite{Takimoto}
\begin{eqnarray} 
&&\hat{\chi}^{\mathrm{s}} \left( q \right) = \frac{\hat{\chi}^0 \left( q \right)}{\hat{1} - \hat{\Gamma}^{\mathrm{s}} \hat{\chi}^0 \left( q \right)}, 
\label{eqn:chis} \\
&&\hat{\chi}^{\mathrm{c}} \left( q \right) = \frac{\hat{\chi}^0 \left( q \right)}{\hat{1} - \hat{\Gamma}^{\mathrm{c}} (\omega_l) \hat{\chi}^0 \left( q \right)},
\label{eqn:chic}
\end{eqnarray}
where
\begin{equation}
(\Gamma^{\mathrm{s}})_{l_{1}l_{2},l_{3}l_{4}}^{\alpha \beta} = \delta_{\alpha \beta} \times \begin{cases}
U, & l_1=l_2=l_3=l_4 \\
U' , & l_1=l_3 \neq l_2=l_4 \\
J, & l_1=l_2 \neq l_3=l_4 \\
J' , & l_1=l_4 \neq l_2=l_3 \\
0 , & \mathrm{otherwise}
\end{cases}
\end{equation}
\begin{equation}
\hat{\Gamma}^{\mathrm{c}} ( \omega_l )= -\hat{C} - 2\hat{V}_{\mathrm{quad}}( \w_l ),
\label{eqn:Gc}
\end{equation}
\begin{equation}
(C)_{l_{1}l_{2},l_{3}l_{4}}^{\alpha \beta} = \delta_{\alpha \beta} \times \begin{cases}
U, & l_1=l_2=l_3=l_4 \\
-U'+2J , & l_1=l_3 \neq l_2=l_4 \\
2U' - J , & l_1=l_2 \neq l_3=l_4 \\
J' , &l_1=l_4 \neq l_2=l_3 \\
0 . & \mathrm{otherwise}
\end{cases}
\end{equation}
$\alpha, \beta ( = A, B)$ is Fe-site.
In this paper, we set $U = U' + 2J$, $J=J'$, and $J = U / 6$.

In the RPA, the enhancement of the spin susceptibility $\hat{\chi}^{\mathrm{s}}$
originates from the intra-orbital Coulomb interaction $U$
and the ``intra-orbital nesting'' of the FSs.
On the other hand, the enhancement of $\hat{\chi}^{\mathrm{c}}$ 
originates from the quadrupole-quadrupole interaction in Eq. (\ref{eqn:Hint})
and the ``inter-orbital nesting'' of the FSs.
The magnetic (orbital) order occurs when $\alpha_{\mathrm{s}(\mathrm{c})}=1$,
where $\alpha_{\mathrm{s}(\mathrm{c})}$ is the the maximum eigenvalue of 
$\hat{\Gamma}^{\mathrm{s}(\mathrm{c})} \hat{\chi}^{(0)} ( \bm{q} , 0)$,
called the spin (charge) Stoner factor.
Here, the magnetic order is realized when $U$ reaches $U_{\mathrm{cr}}= 0.448$ eV.
Also, the orbital order is realized when $g$ reaches $g_{\mathrm{cr}}= 0.132$ eV 
for $U=0$ \cite{Saito-LiFeAs}.
The used interactions are smaller since the energy-scale had been renormalized by
$z=(m^*/m_b)^{-1}$ in the present tight-binding model.

\subsection{Spin-orbit interaction}
In this subsection, we introduce the SOI.
In the presence of SOI for $d$ electron, the $20\times20$ matrix expression 
of the total Hamiltonian at $\bm{k}$ is given by
\begin{eqnarray}
\hat{h}(\bm{k}) &=& \begin{pmatrix}
\hat{h}^0(\bm{k}) + \lambda \hat{l}_z / 2 & \lambda ( \hat{l}_x - i \hat{l}_y) / 2 \\
\lambda ( \hat{l}_x + i \hat{l}_y) / 2 & \hat{h}^0(\bm{k}) - \lambda \hat{l}_z / 2
\end{pmatrix}
 \nonumber \\
&\equiv& \begin{pmatrix}
\hat{h}^{\uparrow \uparrow} (\bm{k}) & \hat{h}^{\uparrow \downarrow} (\bm{k}) \\
\hat{h}^{\downarrow \uparrow} (\bm{k}) & \hat{h}^{\downarrow \downarrow} (\bm{k})
\end{pmatrix} ,
\label{eqn:hk}
\end{eqnarray}
where the first and the second rows (columns) correspond to
$\uparrow$-spin and $\downarrow$-spin.
$\lambda$ is coupling constant of SOI.
The matrix elements for $\hat{\bm{l}}$ for $d$-orbital, 
which is diagonal with respect to site, are given by \cite{Friedel}
\begin{align}
\hat{l}_x &= \begin{pmatrix}
0 & 0 & \sqrt{3} i & 0 & 0 \\
0 & 0 & 0 & i & 0 \\
- \sqrt{3} i & 0 & 0 & 0 & - i \\
0 & - i & 0 & 0 & 0 \\
0 & 0 & i & 0 & 0
\end{pmatrix}, \\
\hat{l}_y &= \begin{pmatrix}
0 & - \sqrt{3} i & 0 & 0 & 0 \\
\sqrt{3} i & 0 & 0 & 0 & - i \\
0 & 0 & 0 & - i & 0 \\
0 & 0 & i & 0 & 0 \\
0 & i & 0 & 0 & 0
\end{pmatrix}, \\
\hat{l}_z &= \begin{pmatrix}
0 & 0 & 0 & 0 & 0 \\
0 & 0 & - i & 0 & 0 \\
0 & i & 0 & 0 & 0 \\
0 & 0 & 0 & 0 & 2i \\
0 & 0 & 0 & -2i & 0
\end{pmatrix},
\end{align}
where the first to fifth rows (columns) correspond to $d$ orbitals
$z^2, xz, yz, xy, x^2-y^2$, respectively.

The red (blue) lines in Fig. \ref{fig:FS} (a) shows the band structure of LiFeAs for 
$\lambda = 0.05$ eV ($\lambda=0$), respectively.
Points A and B are shown in Fig. \ref{fig:FS} (b).
When $\lambda = 0$, the two bands composed of $d_{xz}+d_{yz}$ orbital
near the Fermi level at Z point are degenerated,
however, this degeneracy is lifted by the SOI, which seems to be
observed by recent ARPES measurements in LiFeAs \cite{Borisenko-SOI,Ding-SOI}
as well as in FeSe \cite{FeSe1,FeSe2,FeSe3,FeSe4,FeSe5,FeSe6}.
Figures \ref{fig:FS} (b) and (c) show the FSs of LiFeAs for 
$\lambda = 0$ and $0.05$ eV, respectively.
For this reason, h-FS1 disappears by introducing the SOI.
This model is consistent with the ARPES measurement by Miao {\it et al}.
in the way that only outer band of two $d_{xz/yz}$-orbital hole band 
at Z point crosses the Fermi level \cite{Ding-SOI}.
In Sec. \ref{sec:Borisenko-model}, we also construct the tight-binding model
that gives the ARPES band structure reported by Borisenko {\it et al}
 \cite{Borisenko-SOI}.

Since the energy-scale of the present tight-binding model had been renormalized
by the factor $z=(m^*/m_b)^{-1}$,
we should use a renormalized $\lambda$ in Eq. (\ref{eqn:hk}) in principle.
However, we hereafter use a typical $\lambda$ for Fe ion 
($\lambda=0.05$ eV) \cite{Friedel} without considering the renormalization,
in order to clarify the effect of the SOI.
In fact, we had verified that the obtained gap structures in later sections
are unchanged by using smaller SOI coupling, say $\lambda\sim 0.02$ eV.

\subsection{Gap equation}
In this subsection, we introduce the linearized gap equation
in the presence of the SOI.
In the 3D model, it is difficult to analyze
fine momentum dependence of the SC gap on the FSs
by using a conventional 3D $\k$-meshes.
In this paper, we used $48 \times 16 \ \bm{k}$ points for each FS sheet
in solving the gap equation
in order to obtain the precise gap structure
\cite{Scalapino}.
In the absence of impurities,
the linearized gap equation for the singlet state is given as 
\begin{multline}
\lambda_{\mathrm{E}} \Delta^{L}_{\Sigma \bar{\Sigma}} ( \bm{k} , \e_n ) 
= \frac{\pi T}{(2 \pi )^3} \sum_{\e_m} \sum_{M}^{\mathrm{FS}} \sum_{\Lambda}^{\Uparrow \Downarrow} \int_{\mathrm{FS}M} 
\frac{d \bm{k}'_{\mathrm{FS}M}}{v^{M} ( \bm{k}')} \\
\times V^{LM}_{\Sigma \bar{\Sigma} \Lambda \bar{\Lambda}}( \bm{k} , \bm{k}' , \e_n - \e_m ) 
\frac{\Delta^{M}_{\Lambda \bar{\Lambda}}( \bm{k}' , \e_m )}{| \e_m |},
\label{eqn:Eliasheq}
\end{multline}
where $\lambda_{\mathrm{E}}$ is the eigenvalue of the gap equation.
$\lambda_{\mathrm{E}}$ reaches unity at $T=T_{\mathrm{c}}$.
We solve the gap function for the largest eigenvalue, since it 
corresponds to the pairing state for the highest $T_{\mathrm{c}}$.
$L$ and $M$ denote the FSs, and $\Sigma$ and $\Lambda$ denote the 
pseudo-spin ($\Uparrow , \Downarrow$) in the presence of the SOI.
Note that $\bar{\Sigma} \equiv - \Sigma$.
$\Delta^{L}_{\Sigma \bar{\Sigma}} ( \bm{k} , \e_n )$ is the 
pseudo-spin singlet gap function on the $L$-th FS. 
In Eq. (\ref{eqn:Eliasheq}), we perform the surface integral on the $M$-th FS.
The paring interaction $V$ in Eq. (\ref{eqn:Eliasheq}) 
for the pseudo-spin singlet pair is given by
\begin{multline}
V^{L M}_{\Sigma \bar{\Sigma} \Lambda \bar{\Lambda}} ( \bm{k} , \bm{k}' , \e_n - \e_m) \\
= \sum_{l m m' l'} \sum_{\sigma \lambda \lambda ' \sigma '}
U_{lL}^{\sigma \Sigma *} ( \bm{k} ) U_{l' L'}^{\sigma ' \bar{\Sigma} *} ( - \bm{k} ) \\
\times V_{l m m' l'}^{\sigma \lambda \lambda' \sigma'} (\bm{k} - \bm{k}' , \e_n - \e_m ) \\
\times U_{m' M'}^{\lambda' \bar{\Lambda}} ( - \bm{k}' ) U_{m M}^{\lambda \Lambda} ( \bm{k}' ),
\label{eqn:Vint}
\end{multline}
where $\sigma, \lambda$ mean real spin ($\downarrow , \uparrow$)
and $U_{l L}^{\sigma \Sigma} ( \bm{k} ) = \langle \bm{k} ; l \sigma | \bm{k} ; L \Sigma \rangle$
is the unitary matrix connecting
between the band representation and the orbital one.
Here, the pairing interaction in the orbital basis is given by
\begin{align}
V_{l m m' l'}^{\sigma \lambda \lambda' \sigma'} &= V^{\mathrm{c}}_{l m m' l'} \delta_{\sigma \lambda} \delta_{\sigma' \lambda'}
+V^{\mathrm{s}}_{l m m' l'} \bm{\sigma}_{\sigma \lambda} \cdot \bm{\sigma}_{\sigma' \lambda'} \notag \\
&=
\begin{cases}
V^{\mathrm{c}}_{l m m' l'} + V^{\mathrm{s}}_{l m m' l'}, & \sigma = \lambda = \lambda' = \sigma' , \\
V^{\mathrm{c}}_{l m m' l'} - V^{\mathrm{s}}_{l m m' l'}, & \sigma = \lambda \ne \lambda' = \sigma' , \\
2 V^{\mathrm{s}}_{l m m' l'}, & \sigma = \lambda' \ne \lambda = \sigma' , \\
0 , & \mathrm{otherwise} ,
\end{cases}
\label{eqn:Worb}
\end{align}
\begin{gather}
\hat{V}^{\mathrm{c}} = \frac{1}{2} \hat{\Gamma}^{\mathrm{c}} \hat{\chi}^{\mathrm{c}} \hat{\Gamma}^{\mathrm{c}}, \ \ \ 
\hat{V}^{\mathrm{s}} = \frac{1}{2} \hat{\Gamma}^{\mathrm{s}} \hat{\chi}^{\mathrm{s}} \hat{\Gamma}^{\mathrm{s}}. \label{eqn:Worb-cs}
\end{gather}
In this paper, $\hat{V}^{\xi} (\xi = \mathrm{c,s})$ 
is calculated without the SOI,
since it is verified that the SOI is negligible except at the vicinity of 
the magnetic critical point ($\alpha_s\approx1$).

Now, we introduce the relation between
$U_{l L}^{\sigma \Sigma} ( \bm{k} )$ and $U_{l L}^{\bar{\sigma} \bar{\Sigma}} ( - \bm{k} )$
to preserve the time-reversal symmetry between the states
$|\bm{k};L\Sigma\rangle$ and $|-\bm{k};L\bar{\Sigma}\rangle$.
The $20\times20$ Hamiltonian $\hat{h}(\bm{k})$ is given in Eq. (\ref{eqn:hk}).
Then, $\hat{h} ( - \bm{k} )^{*}$ is given by
\begin{align}
\hat{h} (- \bm{k})^{*} &= \begin{pmatrix}
\hat{h}_0 (- \bm{k})^{*} + \lambda \hat{l}_z^{*} / 2 & \lambda ( \hat{l}_x - i \hat{l}_y)^{*} / 2 \\
\lambda ( \hat{l}_x + i \hat{l}_y)^{*} / 2 & \hat{h}_0(- \bm{k})^{*} - \lambda \hat{l}_z^{*} / 2
\end{pmatrix} \notag \\
&= \begin{pmatrix}
\hat{h}_0 (\bm{k}) - \lambda \hat{l}_z / 2 & - \lambda ( \hat{l}_x + i \hat{l}_y) / 2 \\
- \lambda ( \hat{l}_x - i \hat{l}_y) / 2 & \hat{h}_0 (\bm{k}) + \lambda \hat{l}_z / 2
\end{pmatrix} \notag \\
&= \begin{pmatrix}
\hat{h}^{\downarrow \downarrow} (\bm{k}) & - \hat{h}^{\downarrow \uparrow} (\bm{k}) \\
- \hat{h}^{\uparrow \downarrow} (\bm{k}) & \hat{h}^{\uparrow \uparrow} (\bm{k})
\end{pmatrix} . \label{eq:H(-k)}
\end{align}
Here, we express the unitary matrix as
$\displaystyle \hat{U} ( \bm{k} ) \equiv \bigl( \begin{smallmatrix}
\hat{U}^{\mathrm{\uparrow \Uparrow}} (\bm{k}) & \hat{U}^{\mathrm{\uparrow \Downarrow}} (\bm{k}) \\
\hat{U}^{\mathrm{\downarrow \Uparrow}} (\bm{k}) & \hat{U}^{\mathrm{\downarrow \Downarrow}} (\bm{k})
\end{smallmatrix} \bigl)$ that diagonalize $\hat{h} ( \bm{k} )$.
That is,
\begin{equation}
\hat{U} ( \bm{k} )^{\dagger} \hat{h} ( \bm{k} ) \hat{U} ( \bm{k} )
=\begin{pmatrix}
\hat{E} ( \bm{k} ) & \hat{0} \\ \hat{0} & \hat{E} ( \bm{k} )
\end{pmatrix} , \label{eq:diagonalize}
\end{equation}
where $\hat{E} ( \bm{k} )$ is diagonal.
By using Eqs. (\ref{eq:H(-k)}) and (\ref{eq:diagonalize}), we obtain that 
\begin{multline}
\begin{pmatrix} \hat{U}^{\downarrow \Downarrow} (\bm{k})^{\dagger} & - \hat{U}^{\uparrow \Downarrow} (\bm{k})^{\dagger} \\
- \hat{U}^{\downarrow \Uparrow} (\bm{k})^{\dagger} & \hat{U}^{\uparrow \Uparrow} (\bm{k})^{\dagger}
\end{pmatrix}
\hat{h} (- \bm{k})^{*} \\
\times
\begin{pmatrix} \hat{U}^{\downarrow \Downarrow} (\bm{k}) & - \hat{U}^{\downarrow \Uparrow} (\bm{k}) \\
- \hat{U}^{\uparrow \Downarrow} (\bm{k}) & \hat{U}^{\uparrow \Uparrow} (\bm{k}) \end{pmatrix}
=
\begin{pmatrix}
\hat{E} ( \bm{k} ) & \hat{0} \\ \hat{0} & \hat{E} ( \bm{k} )
\end{pmatrix} .
\end{multline}
Therefore, $\hat{U} ( - \bm{k} )$ is related to $\hat{U}(\bm{k})$ as
\begin{equation}
\hat{U} ( - \bm{k} ) = \begin{pmatrix}
\hat{U}^{\downarrow \Downarrow} (\bm{k})^{*} & - \hat{U}^{\downarrow \Uparrow} (\bm{k})^{*} \\
- \hat{U}^{\uparrow \Downarrow} (\bm{k})^{*} & \hat{U}^{\uparrow \Uparrow} (\bm{k})^{*} \end{pmatrix} .
\end{equation}
That is, the relations between
$U_{l L}^{\sigma \Sigma} ( \bm{k} )$ and $U_{l L}^{\bar{\sigma} \bar{\Sigma}} ( - \bm{k} )$ 
are summarized as
\begin{align}
U_{l L}^{\uparrow \Uparrow} ( - \bm{k} ) &= U_{l L}^{\downarrow \Downarrow} ( \bm{k} )^{*}, \notag \\
U_{l L}^{\downarrow \Downarrow} ( - \bm{k} ) &= U_{l L}^{\uparrow \Uparrow} ( \bm{k} )^{*}, \notag \\
U_{l L}^{\uparrow \Downarrow} ( - \bm{k} ) &= - U_{l L}^{\downarrow \Uparrow} ( \bm{k} )^{*}, \notag \\
U_{l L}^{\downarrow \Uparrow} ( - \bm{k} ) &= - U_{l L}^{\uparrow \Downarrow} ( \bm{k} )^{*}.
\end{align}
These relations are necessary to obtain physical gap functions.

Next, we transform the gap equation (\ref{eqn:Eliasheq}) to 
a more convenient expression.
By setting $\Sigma = \Uparrow \ (\bar{\Sigma} = \Downarrow)$ 
and taking the summation of $\Lambda$, gap equation is given as
\begin{multline}
\lambda_{\mathrm{E}} \Delta^{L}_{\mathrm{ \Uparrow \Downarrow}} ( \bm{k} , \e_n ) 
= \frac{\pi T}{(2 \pi )^3} \sum_{\e_m} \sum_{M}^{\mathrm{FS}} \int_{\mathrm{FS}M} 
\frac{d \bm{k}'_{\mathrm{FS}M}}{v^{M} ( \bm{k}')} \\
\times \left\{ V^{LM}_{\Uparrow \Downarrow \Uparrow \Downarrow} ( \bm{k} , \bm{k}' , \e_n - \e_m ) 
\frac{\Delta^{M}_{\Uparrow \Downarrow}( \bm{k}' , \e_m )}{| \e_m |} \right. \\
+ \left. V^{LM}_{\Uparrow \Downarrow \Downarrow \Uparrow} ( \bm{k} , \bm{k}' , \e_n - \e_m ) 
\frac{\Delta^{M}_{\Downarrow \Uparrow}( \bm{k}' , \e_m )}{| \e_m |} \right\}
\end{multline}
Since the relation $\Delta^{L}_{\Uparrow \Downarrow} ( k ) 
= - \Delta^{L}_{\Downarrow \Uparrow} (k)$ is satisfied, 
the gap equation for the singlet pairing state is given as
\begin{multline}
\lambda_{\mathrm{E}} \Delta^{L} ( \bm{k} , \e_n ) 
= \frac{\pi T}{(2 \pi )^3} \sum_{\e_m} \sum_{M}^{\mathrm{FS}} \int_{\mathrm{FS}M} 
\frac{d \bm{k}'_{\mathrm{FS}M}}{v^{M} ( \bm{k}')} \\
\times \left\{ V^{LM}_{\Uparrow \Downarrow \Uparrow \Downarrow} ( \bm{k} , \bm{k}' , \e_n - \e_m ) \right. \\
\left.- V^{LM}_{\Uparrow \Downarrow \Downarrow \Uparrow} ( \bm{k} , \bm{k}' , \e_n - \e_m ) \right\}
\frac{\Delta^{M}( \bm{k}' , \e_m )}{| \e_m |}
\label{eqn:EliashSOI}
\end{multline}
where $\Delta^{L} ( \bm{k} , \e_n ) \equiv 
\Delta^{L}_{\Uparrow \Downarrow} ( \bm{k} , \e_n )$.

\section{Superconducting gap for the SOI-induced bandstructure (I)}
\label{sec:RPA}
In this section, we numerically analyze the 
linearized Eliashberg equation, Eq. (\ref{eqn:EliashSOI}),
on the basis of the SOI-induced bandstructure in Fig. \ref{fig:FS} (a),
which we call the ``bandstructure (I)''.
We show the obtained 3D gap function $\Delta^L (\theta , k_z)$
on the FS sheet $L$.
Here, we divide the variables $\theta = [0 , 2 \pi ]$ and $k_z = [- \pi, \pi ]$ into 48 and 16 meshes, respectively,
and use 512 Matsubara frequencies.
The pairing interaction in Eq. (\ref{eqn:Vint}) is given by the RPA.
%

\subsection{Orbital-fluctuation-mediated $s_{++}$-wave state}
\label{sec:s++gap}

We first discuss the $s_{++}$-wave state that is 
realized by orbital-fluctuations in the presence of the SOI:
Figure \ref{fig:gap1} shows the obtained gap functions
in the case of $g = 0.129$ eV and $U = 0$ ($\alpha_c = 0.98$) 
in the $k_z = \pi$ plane.
The overall gap structure is essentially unchanged for $\a_c=0.90\sim0.98$
\cite{Saito-LiFeAs}.
As for the hole pockets, the gap functions on the h-FS2 
composed of $(d_{xz},d_{yz})$ orbitals are the largest in magnitude,
while the gap on the h-FS3 composed of the $d_{xy}$ orbital is the smallest.
This result is quantitatively consistent with the experimental reports
in Refs. \cite{Borisenko-LiFeAs,Ding-SOI}:
Experimental gap functions in Ref. \cite{Borisenko-LiFeAs}
are shown by dots in Fig. \ref{fig:gap1}.
This result is also similar to our previous result without SOI,
regardless of the presence or absence of the h-FS1 \cite{Saito-LiFeAs}.
Here, we adjust the magnitude of gap functions 
given by solving the linearized gap equation.

\begin{figure}[!htb]
\includegraphics[width=0.9\linewidth]{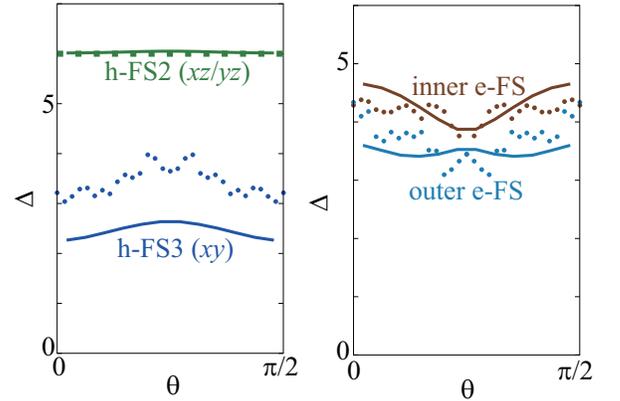}
\caption{
(Color online)
Obtained gap functions for $U=0$ and $g=0.129$ eV ($s_{++}$-wave state)
in the $k_z=\pi$-plane for the SOI-induced bandstructure (I).
The dots represent the experimental data  (in unit meV)
given by the ARPES measurement in Ref. \cite{Borisenko-LiFeAs}.
}
\label{fig:gap1}
\end{figure}

As for the electron pockets,
the gap of the inner e-FS is larger than that of the outer e-FS,
and has the local maxima at $\theta = 0$ and $\pi/2$
and the minima at $\theta = \pi / 4$.
This result is also consistent with the experimental data  
\cite{Borisenko-LiFeAs}, and it is 
very similar to our previous result without SOI \cite{Saito-LiFeAs}.
The obtained gap structure is almost independent of $k_z$
except for the three dimensional pocket h-FS2.
These obtained results are essentially independent of the strength of $g$.

Next, we discuss the origin of the orbital and FS dependencies of 
the gap functions.
For finite $g$, orbital-fluctuations develop due to the good 
inter orbital nesting between h-FS2 (orbital 2,3) and e-FSs (orbital 4).
For this reason, the maximum gap is realized on h-FS2 ($\Delta^{\mathrm{h}}_2$),
and inner e-FS ($\Delta^{\mathrm{e}}_{\mathrm{out}}$) at $\theta = 0$ and $\pi/2$.
This result is not changed by considering the SOI.
Therefore, the experimentally observed gap functions are understood 
based on the orbital-fluctuation theory.

\subsection{Spin-fluctuation-mediated $s_{\pm}$-wave state}
\label{sec:s+-gap}
Next, we discuss the $s_{\pm}$-wave state realized by spin-fluctuations:
Figure \ref{fig:gap2} shows the obtained gap structure
in the case of $g = 0$ and $U = 0.439$ eV ($\alpha_s = 0.98$) 
in the $k_z = \pi$ plane.
The overall gap structure is essentially unchanged for $\a_s=0.90\sim0.98$
\cite{Saito-LiFeAs}.
The gap function on each FS is almost independent of $k_z$.
The obtained gap structure is essentially independent of the value of $U$.
The obtained very small gap on the h-FS2 is
totally opposite to the experimental data shown by dots.

\begin{figure}[!htb]
\includegraphics[width=0.9\linewidth]{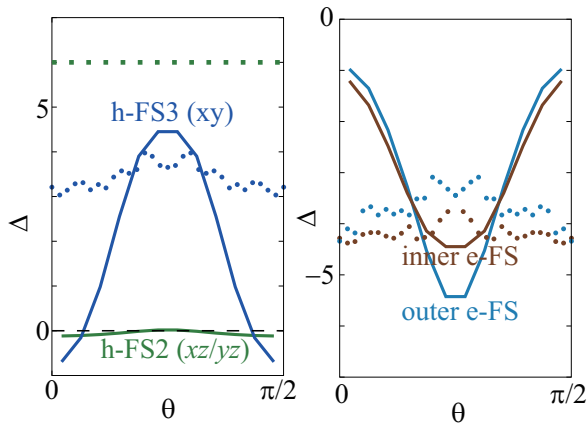}
\caption{
(Color online)
Obtained gap functions for $U=0.439$ eV and $g=0$ ($s_{\pm}$-wave state)
in the $k_z=\pi$-plane.
The dots represent the experimental data  (in unit meV) 
\cite{Borisenko-LiFeAs}.
}
\label{fig:gap2}
\end{figure}

In addition, the obtained $\theta$ dependence of the gap on the e-FSs and h-FS3
is very different from the experimental data.
Both the gap on the h-FS3 and e-FSs show the maximum values at $\theta = \pi/4$,
since they are connected by the wave vector of spin-fluctuations $\bm{Q} \sim ( \pi , 0),(0, \pi )$.
Moreover, the gap function on h-FS3 has eight nodes,
which is inconsistent with experiments.
These results are very similar to our previous results without SOI 
\cite{Saito-LiFeAs}.
Note that these eight nodes on the hole-FSs
disappear in the case of $J/U=0.4$
($U'\equiv U-2J=0.2U$) in accordance with the previous study in
Ref. \cite{Hirschfeld-LiFeAs}.

\subsection{Coexistence of orbital- and spin-fluctuations}
Now, we discuss the superconducting state when the orbital- and spin-fluctuations coexist.
In the previous results without SOI,
coexistence of orbital- and spin-fluctuations leads to an exotic $s$-wave state,
which has sign reversal within h-FSs: $\Delta_{2}^{\mathrm{h}} \Delta_{3}^{\mathrm{h}} < 0$ \cite{Saito-LiFeAs}.
Here, we show that this "hole-$s_{\pm}$-state" is also realized 
in the presence of the SOI.

Figure \ref{fig:gap3} (a) shows the obtained gap functions 
in the case of $g = 0.125$ eV and $U = 0.200$ eV.
The realized Stoner factors are $\alpha_c = 0.98$ and $\alpha_s = 0.45$.
In this case, the orbital-fluctuations are much larger than the spin-fluctuations,
and therefore we obtain the $s_{++}$-wave state.
Except for h-FS3, the obtained gap structures are similar to those of 
the "pure $s_{++}$ state" without $U$ in Fig. \ref{fig:gap1}.
Due to the moderate spin-fluctuations on the $d_{xy}$ orbital,
the anisotropy of $\Delta_3^{\mathrm{h}}$ is enlarged.

\begin{figure}[!htb]
\includegraphics[width=0.9\linewidth]{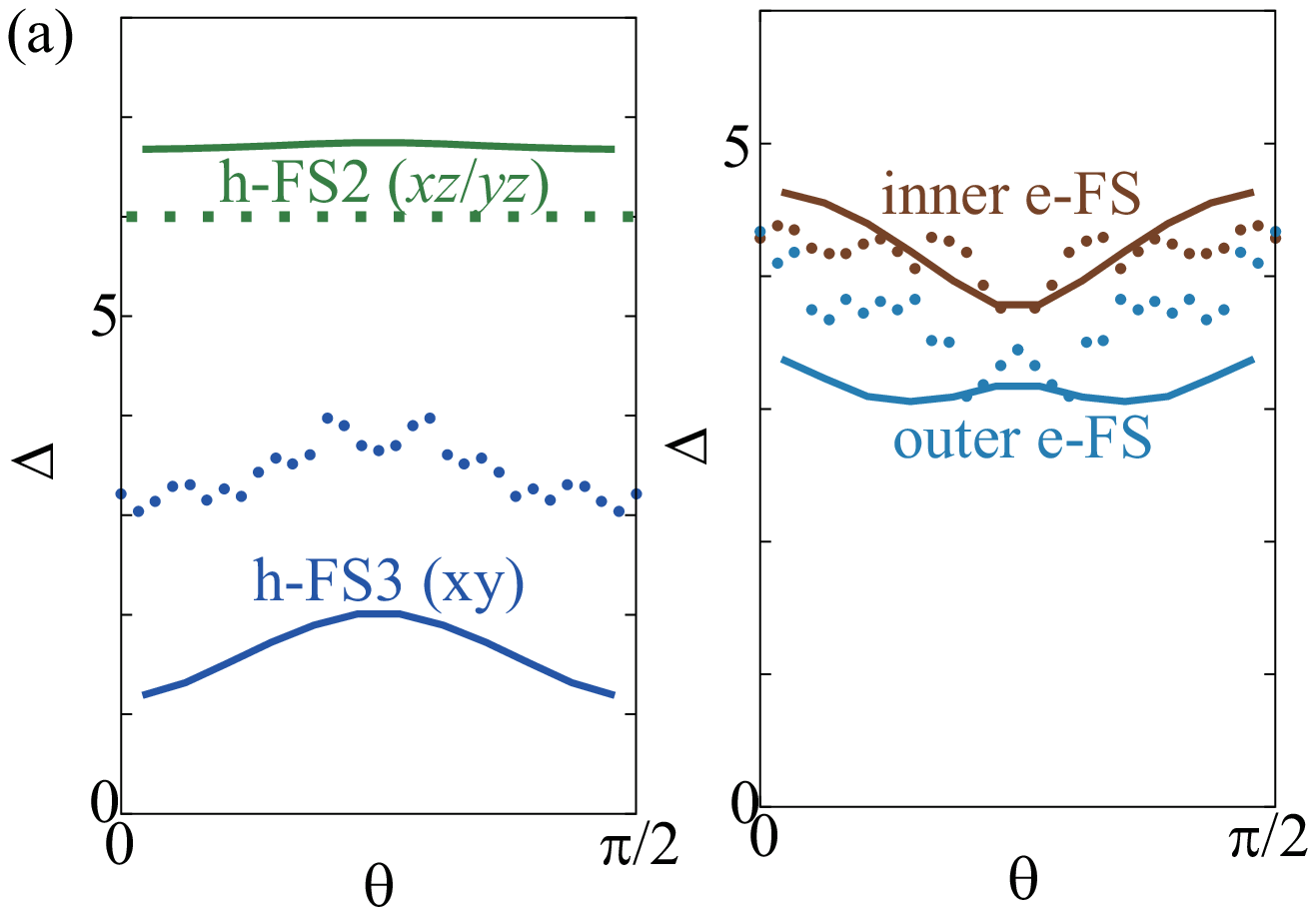}
\includegraphics[width=0.9\linewidth]{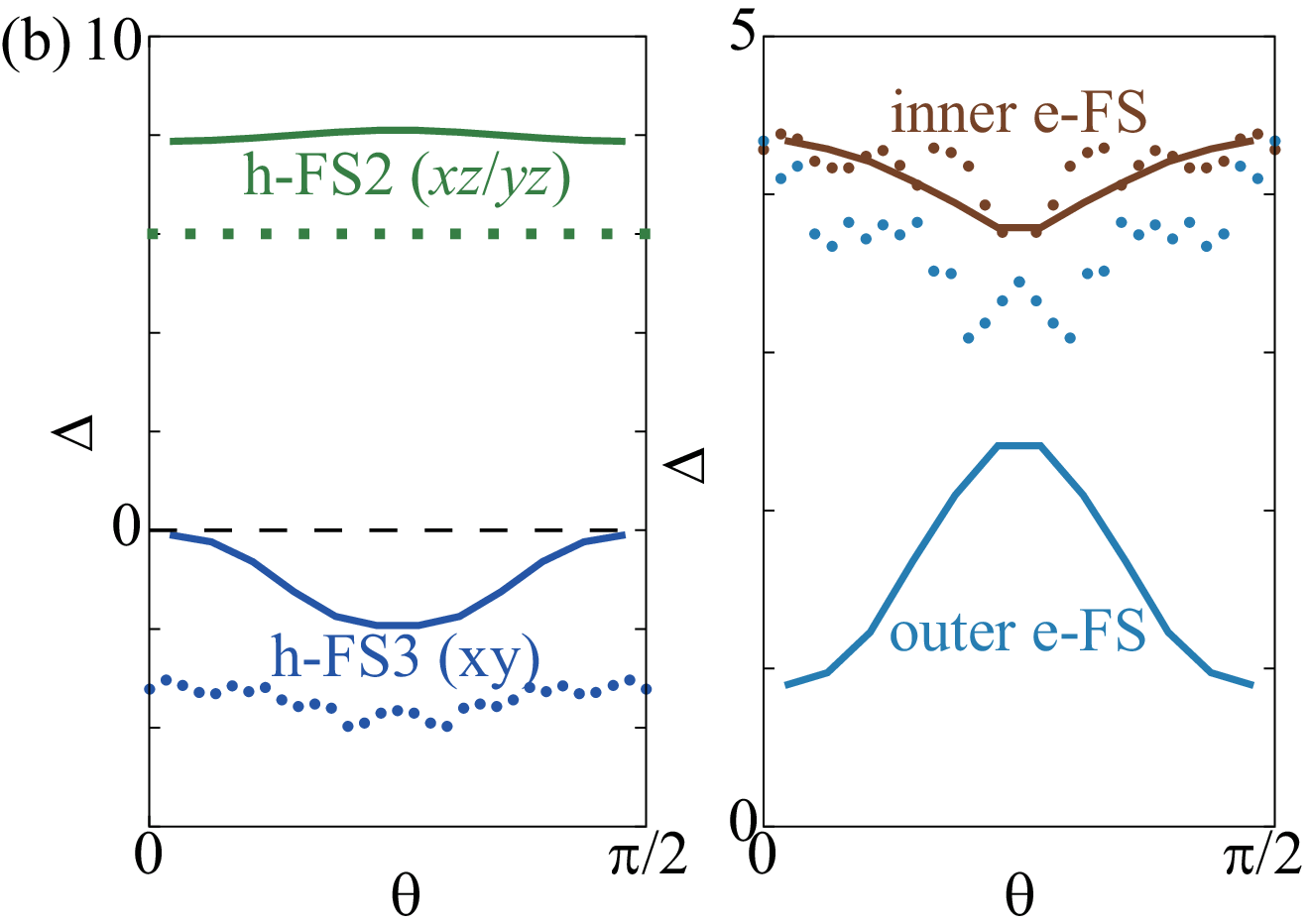}
\caption{
(Color online)
(a) Obtained gap functions for $U=0.200$ eV and $g=0.125$ ($s_{++}$-wave state)
in the $k_z=\pi$-plane.
The dots represent the experimental data (in unit meV) \cite{Borisenko-LiFeAs}.
(b) Obtained gap functions for $U=0.380$ eV and $g=0.122$ 
(hole-$s_{\pm}$-wave state) in the $k_z=\pi$-plane.
}
\label{fig:gap3}
\end{figure}

If we increase the value of $U$ further,
we obtain the hole-$s_{\pm}$-state with the sign reversal within hole pockets.
Figure \ref{fig:gap3} (b) shows the obtained gap functions
in the case of $g = 0.122$ eV and $U = 0.380$ eV 
($\alpha_c = 0.98$ and $\alpha_s = 0.85$).
Here, only $\Delta_3^{\mathrm{h}}$ is negative.
In this hole-$s_{\pm}$-wave state,
the gap structure of $\Delta^{\mathrm{h}}_2$ and
that of the e-FSs ($\Delta^{\mathrm{e}}$)
are qualitatively similar to those in the $s_{++}$-wave state 
in Fig. \ref{fig:gap1}.
On the other hand, $\Delta^{\mathrm{h}}_3$ becomes very anisotropic,
similarly to $\Delta^{\mathrm{h}}_3$ in the $s_{\pm}$-wave state 
in Fig. \ref{fig:gap2}.

In the hole-$s_{\pm}$-wave state given by the
coexistence of orbital- and spin-fluctuations,
$\Delta^{\mathrm{h}}_2 \cdot \Delta^{\mathrm{e}}$ 
is positive due to orbital-fluctuations,
whereas $\Delta^{\mathrm{h}}_3 \cdot \Delta^{\mathrm{e}}$ is negative 
due to spin-fluctuations.
Although the gap on the tiny hole-pocket in the hole-$s_{\pm}$-wave state
takes the largest value,
the overall gap structure is not consistent with ARPES measurement 
in Ref. \cite{Borisenko-LiFeAs},
The present hole-$s_{\pm}$-wave
due to cooperation of orbital- and spin-fluctuation
would be realized in other iron based superconductors.
For instance, the hole-$s_{\pm}$-wave state is first discussed 
in Ba$_{1-x}$K$_x$Fe$_2$As$_2$ based on the thermal conductivity 
and penetration depth measurements \cite{Watanabe-shpm},
in addition to the recent ARPES study \cite{Zhang-shpm}.

\begin{figure}[!htb]
\includegraphics[width=0.99\linewidth]{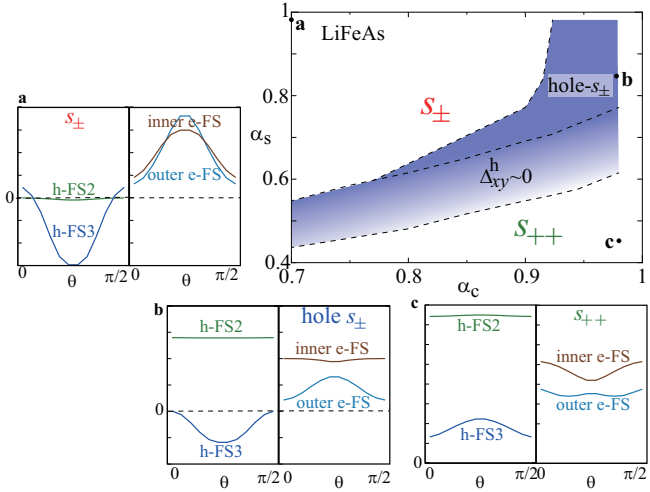}
\caption{
(Color online)
$\alpha_c$-$\alpha_s$ phase diagram of the gap structure in LiFeAs
obtained for the bandstructure (I) in Sec. \ref{sec:RPA}.
Similar phase diagram is obtained for the bandstructure (II).
The gap functions at each point {\bf a} $\sim$ {\bf c} are shown.
}
\label{fig:PB}
\end{figure}

Figure \ref{fig:PB} shows the obtained $\alpha_c$-$\alpha_s$ phase diagram 
of the gap structure in LiFeAs for $\lambda = 0.05$ eV
(bandstructure (I)).
As expected, the $s_{\pm}$-state ($s_{++}$-state) is realized 
for a wide region of $\alpha_c < \alpha_s$ ($\alpha_c > \alpha_s$).
When both $\alpha_c$ and $\alpha_s$ are close to unity, 
we obtain the hole-$s_{\pm}$ state gap in a wide region.
The gap structure at each point {\bf a} $\sim$ {\bf c} is shown in the figure.
In the region ''$\Delta_3^{\mathrm{h}} \sim 0$'',
the obtained $\Delta_3^{\mathrm{h}} ( \theta )$ is very small and has line nodes,
whereas the gaps on the other FSs are large and their signs are the same.
 Thus, nearly $s_{++}$-wave state is realized.
Thus, various types of $s$-wave gap structures are realized 
due to the coexistence of orbital- and spin-fluctuations.

Finally, we comment that the hole-$s_\pm$ state is also realized
in the presence of strong repulsive pairing interaction within the h-FSs,
as discussed in Refs. \cite{Chubukov-spmh,Ahn}.
The authors in Ref. \cite{Yin} also discussed the
hole-$s_\pm$ state due to strong repulsive interaction
between $d_{xy}$ orbital in h-FS3 and $d_{xz,yz}$ orbitals in e-FSs \cite{Yin}.
In the latter scenario, the gaps on e-FSs are nodal in the presence of the SOI.
In contrast, the  hole-$s_\pm$ state in the present paper
is realized by the cooperation between the 
``attractive interaction among $(d_{xz},d_{yz})$- and $d_{xy}$-orbitals''
and ``repulsive interaction on the $d_{xy}$-orbital''.

\section{Another SOI-induced bandstructure and gap function for LiFeAs}
\label{sec:Borisenko-model}

\subsection{SOI-induced bandstructure (II)}
\label{sec:model-II}

\begin{table*}
\caption{Additional hopping integrals $\Delta t_{l,l} (\bm{R})$ 
to change the $d_l$-orbital level by $\Delta E_{l}$ only around the Z point,
without changing the $d_l$-level at M point in the unfolded Brillouin zone.
$s_l=+1$ ($-1$) for $l=z^2,xy,x^2-y^2$ ($xz,yz$).
}
\begin{tabular}{c|ccccc}
\hline
{\small $\bm{R}$}
        & [0,0,0]          & [1,0,0]            & [1,1,0]           & [2,0,0]           &[2,1,0]              \\ 
\hline
  $\Delta t_{l,l}(\bm{R})$ &$(1/16) \Delta E_{l}$ \ & \ $-(3s_l/64) \Delta E_{l}$ \ & \ $(1/32) \Delta E_{l}$ \ & \ $(1/64) \Delta E_{l}$\ & \ $ -(s_l/128) \Delta E_{l}$ \\
\hline
\end{tabular}
\begin{tabular}{c|ccccc}
\hline
{\small $\bm{R}$}
        & [0,0,1]          & [1,0,1]            & [1,1,1]           & [2,0,1]           &[2,1,1]              \\ 
\hline
  $\Delta t_{l,l}(\bm{R})$ &$-(1/32) \Delta E_{l}$\ & \ $(3s_l/128) \Delta E_{l}$ \ & \ $-(1/64) \Delta E_{l}$ \ & \ $-(1/128) \Delta E_{l}$ \ & \ $(s_l/256) \Delta E_{l}$ \\
\hline
\end{tabular}

\label{hopping}
\end{table*}

Recently, the 3D bandstructure in LiFeAs
had been carefully measured by ARPES study in Ref. \cite{Borisenko-SOI}.
The authors observed the strongly dispersing $d_{z^2}+d_{xy}$-orbital band 
crossing the Fermi level along $\Gamma$-Z line, with the minimum at Z point.
Although this dispersion is absent in bandstructure (I),
it can be reproduced by lowering the 
$d_{xy}$-orbital level below the Fermi level at the Z point
without changing $d_{xy}$-orbital h-FS3.
To change the $d_l$-orbital level by $\Delta E_{l}$ 
($l=xz,yz,xy,x^2-y^2,z^2$) only around Z point, 
we introduce the additional intra-orbital hopping integrals shown in TABLE I.
In this procedure,
the $d_l$-level at M point in the unfolded Brillouin zone is unchanged.
Figure \ref{fig:newband-Li} (a) shows the bandstructure 
for $\Delta E_{xy} = - 0.6$ eV and $\Delta E_{xz} = \Delta E_{yz}=0.01$ eV
in the absence of the SOI.
Due to large negative $\Delta E_{xy}$, 
the strongly dispersing band along the $\Gamma$-Z line,
shown by brown line, is reproduced.
The energy of this dispersion is +0.21 eV (-0.51 eV) at $\Gamma$ (Z) point.
The Fermi level is set as zero.
The FSs in the $k_y=0$ plane are shown in Fig. \ref{fig:newband-Li} (b):
After introducing large negative $\Delta E_{xy}$,
$d_{xy}$-level is lower than $d_{xz/yz}$-level at Z-point,
and the former level (latter level) corresponds to the energy of
the hole-band for h-FS1 (h-FS2) at Z-point.
As a result, the degeneracy of the hole-bands for h-FS1,2 at Z point is lifted,
and only h-FS2 remains in Figs. \ref{fig:FS} (a)-(b)
whereas the hole-band for h-FS1 is below $-0.1$ eV.
Thus, the number of the tiny 3D hole-pockets
is reduced to one.

\begin{figure}[!htb]
\includegraphics[width=0.90\linewidth]{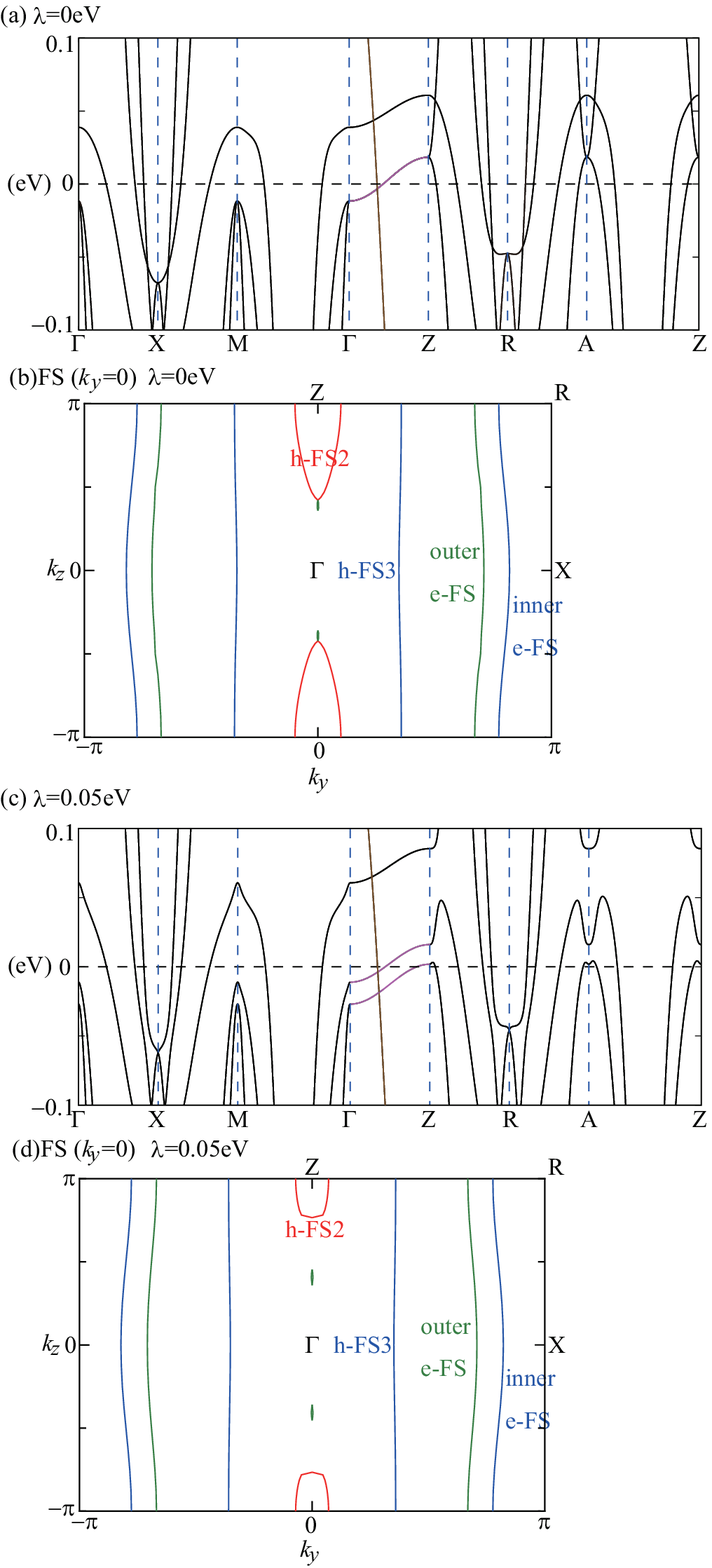}
\caption{(Color online) 
(a) The band structure of the ten-orbital model
for $\Delta E_{xy} = -0.6$ eV and $\Delta E_{xz} = \Delta E_{yz}=0.01$ eV
without the SOI ($\lambda = 0$) and
(b) the FSs in the $k_y = 0$ plane for $\lambda=0$.
Due to large negative $\Delta E_{xy}$, h-FS1 around Z point disappears.
(c) The ``SOI-induced bandstructure (II)'' with $\lambda = 0.05$ eV, and
(d) the corresponding FSs in the $k_y = 0$ plane.
The size of h-FS2 is reduced by introducing the SOI.
}
\label{fig:newband-Li}
\end{figure}

In the next stage, we introduce the SOI ($\lambda=0.05$ eV):
The realized SOI-induced bandstructure is shown in 
Fig. \ref{fig:newband-Li} (c),
which we call the  ``SOI-induced bandstructure (II)''.
Due to the SOI, the degenerated $d_{xz/yz}$-orbital hole-bands
along the $\Gamma$-Z line, is split into two bands,
as shown by purple lines in Fig. \ref{fig:newband-Li} (c).
As a result, the size of h-FS2
is reduced as explained in Ref. \cite{Borisenko-SOI} in detail.
(In contrast, the higher band forms the tiny hole-pocket
in the SOI-induced bandstructure (I) in Sec. \ref{sec:RPA}.)

The obtained SOI-induced FSs in the $k_y=0$ plane are shown in 
Fig. \ref{fig:newband-Li} (d):
In addition to the tiny and large hole-pockets h-FS2 and h-FS3,
there are fine h-FSs due to the Dirac dispersion at 
$| k_z | \sim 0.5\pi$ on the $k_z$-line.
The overall  bandstructure near the Fermi level as well as the
SOI-induced change in the topology of the FSs reported in 
Ref. \cite{Borisenko-SOI}
are qualitatively reproduced in the present bandstructure (II).

\subsection{Superconducting gap for the bandstructure (II)}
\label{sec:RPA-II}

Next, we analyze the linearized Eliashberg equation, Eq. (\ref{eqn:EliashSOI}),
for the SOI-induced bandstructure (II):
The FSs in the $k_z=\pi$ plane are shown in Fig. \ref{fig:newband-Li2} (a).
We stress that the orbital character of h-FS2 in Fig. \ref{fig:newband-Li2} (a)
is essentially the same as that of h-FS2 in Fig. \ref{fig:FS} (c).
This fact indicates that similar gap structures are obtained 
based on both bandstructures (I) and (II).

\begin{figure}[!htb]
\includegraphics[width=0.85\linewidth]{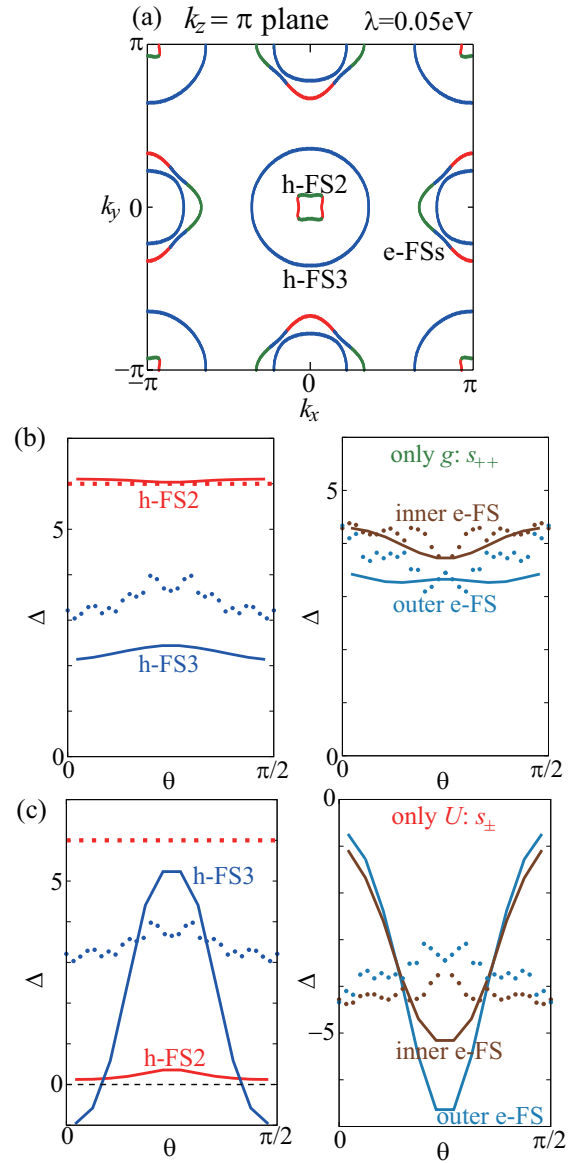}
\caption{(Color online) 
(a) FSs in the  bandstructure (II) with $\lambda=0.05$eV 
in the $k_z=\pi$ plane.
(b)-(c) The obtained gap functions in bandstructure (II) 
in the $k_z = \pi$-plane
for (b) $g = 0.127$ eV and $U = 0$ ($s_{++}$-wave state), and
(c) $U = 0.434$ eV and $g = 0$ ($s_{\pm}$-wave state).
The dots represent the experimental date given by 
ARPES study (in unit meV) \cite{Borisenko-LiFeAs}.
}
\label{fig:newband-Li2}
\end{figure}

The pairing interaction is given by Eqs. (\ref{eqn:Vint}) and (\ref{eqn:Worb})
in the present model with $\Delta E_l$.
We first discuss the $s_{++}$-wave state realized by orbital-fluctuations:
Figure \ref{fig:newband-Li2} (b) shows the obtained gap functions
in the case of $g = 0.127$ eV and $U = 0$ ($\alpha_c = 0.98$) in the $k_z = \pi$ plane.
As for the hole pockets, the gap functions on the h-FS2 composed of $(d_{xz},d_{yz})$ orbitals are the largest,
while the gap on the h-FS3 composed of the $d_{xy}$ orbital is the smallest.
These results are quantitatively consistent with the experimental data \cite{Borisenko-LiFeAs} shown by dots.
As for the electron pockets,
the gap of inner e-FS is larger than that of outer e-FS,
and their $\theta$ dependences are qualitatively
consistent with experimental data in Ref. \cite{Borisenko-LiFeAs}.
The obtained overall gap structure is
similar to that for the bandstructure (I) shown in Sec. \ref{sec:s++gap},
regardless of significantly different bandstructures.

Next, we discuss the $s_{\pm}$-wave state realized by spin-fluctuations:
Figure \ref{fig:newband-Li2} (c) shows the obtained gap structure
in the case of $U = 0.434$ eV and $g = 0$ ($\alpha_s = 0.98$)
in the $k_z = \pi$ plane.
The obtained very small gap functions on the h-FS2
is opposite to the experimental data \cite{Borisenko-LiFeAs} 
shown by dots.
In addition, the obtained $\theta$ dependences of the gaps 
on the e-FSs and h-FS3
are very different from the experimental data:
Both the gaps on the h-FS3 and e-FSs show the maximum values at 
$\theta = \pi/4$ since they are connected by the wave vector of 
spin-fluctuations $\bm{Q} \sim ( \pi , 0),(0, \pi )$.
These results are similar to those
for the bandstructure (I) shown in Sec. \ref{sec:s+-gap}.

\section{Summary}
In this paper, 
we studied the pairing mechanism in LiFeAs by taking the SOI into account.
For this purpose, we constructed possible two 
realistic 3D tight-binding models with SOI,
according the experimental band structures \cite{Borisenko-SOI,Ding-SOI}.
In both SOI-induced bandstructures,
one of the 3D tiny hole-pockets around the Z point disappears
due to the SOI.
The orbital character of the realized tiny single hole-pocket 
is same in both models.
Experimentally, the superconducting gap 
takes the largest value on the tiny single hole-pocket.

Next, we analyzed the 3D gap structure 
in order to determine the correct pairing mechanism.
In the orbital-fluctuation scenario for the $s_{++}$-wave state,
the pairing interaction is given by the orbital-fluctuations 
driven by the inter-orbital nesting 
between ($d_{xz/yz}$, $d_{xy}$) orbitals.
The obtained gap function on the tiny hole-pocket is the largest,
by reflecting the better inter-orbital nesting
between the tiny hole-pocket and the electron-pockets.
The obtained numerical results shown in 
Fig. \ref{fig:gap1} and Fig. \ref{fig:newband-Li2} (b)
are quantitatively consistent with the experimental gap structure in LiFeAs.
In contrast,
in the spin-fluctuation scenario for the $s_\pm$-wave state,
the obtained gap function on the tiny hole-pocket is the smallest,
by reflecting the bad intra-orbital nesting
between the tiny hole-pocket and the electron-pockets.
Thus, the experimental gap structure in LiFeAs 
is unable to be explained by this scenario,
as demonstrated in Fig. \ref{fig:gap2} and in Fig. \ref{fig:newband-Li2} (c).

Therefore, it is concluded that
the inter-orbital ($d_{xz/yz}$-$d_{xy}$) interactions, 
which are given by orbital-fluctuations,
cause the superconductivity in LiFeAs.
The present study reinforces the results obtained in
our previous study for LiFeAs without the SOI \cite{Saito-LiFeAs}.
When both the orbital- and spin-fluctuations develop comparably,
we frequently obtain the ``hole-$s_{\pm}$-state'', in which the 
gap structure has the ``sign reversal between h-FSs''.
This interesting gap structure had been discussed in 
Ba$_{1-x}$K$_x$Fe$_2$As$_2$ based on the thermal conductivity and 
penetration depth measurement \cite{Watanabe-shpm}.
In Appendix A, we show that the hole-$s_{\pm}$-wave gap structure 
can be obtained in the model of Ba122 system.

\acknowledgements
We are grateful to S.V. Borisenko, Y. Matsuda, A. Fujimori, S. Shin, 
and T. Shimojima for valuable discussions.
This study has been supported by Grants-in-Aid for Scientific 
Research from MEXT of Japan.

\appendix
\section{Hole-$s_{\pm}$-wave state due to coexistence of
orbital- and spin-fluctuations in Ba122 systems}

\begin{figure}[!htb]
\includegraphics[width=0.99\linewidth]{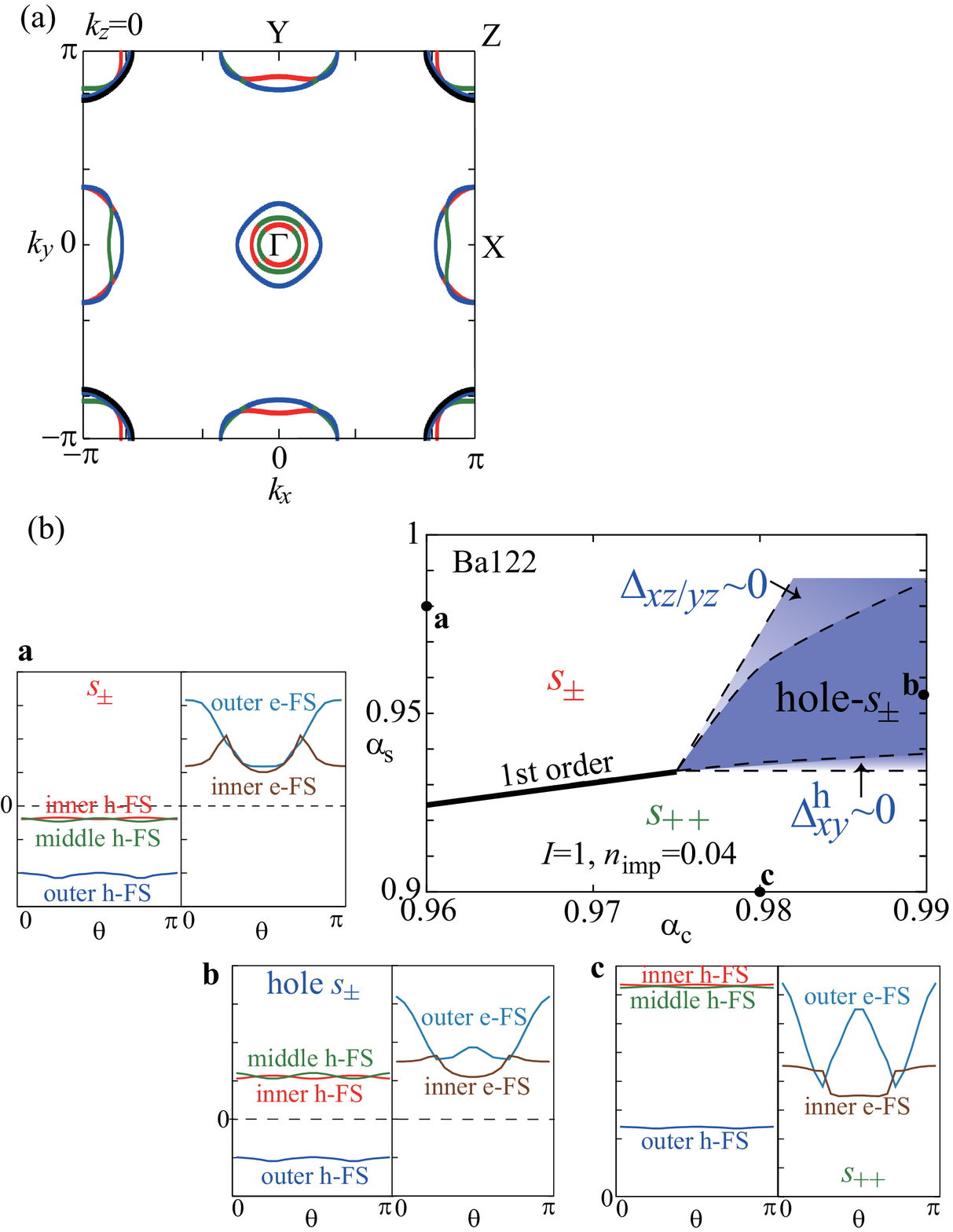}
\includegraphics[width=0.8\linewidth]{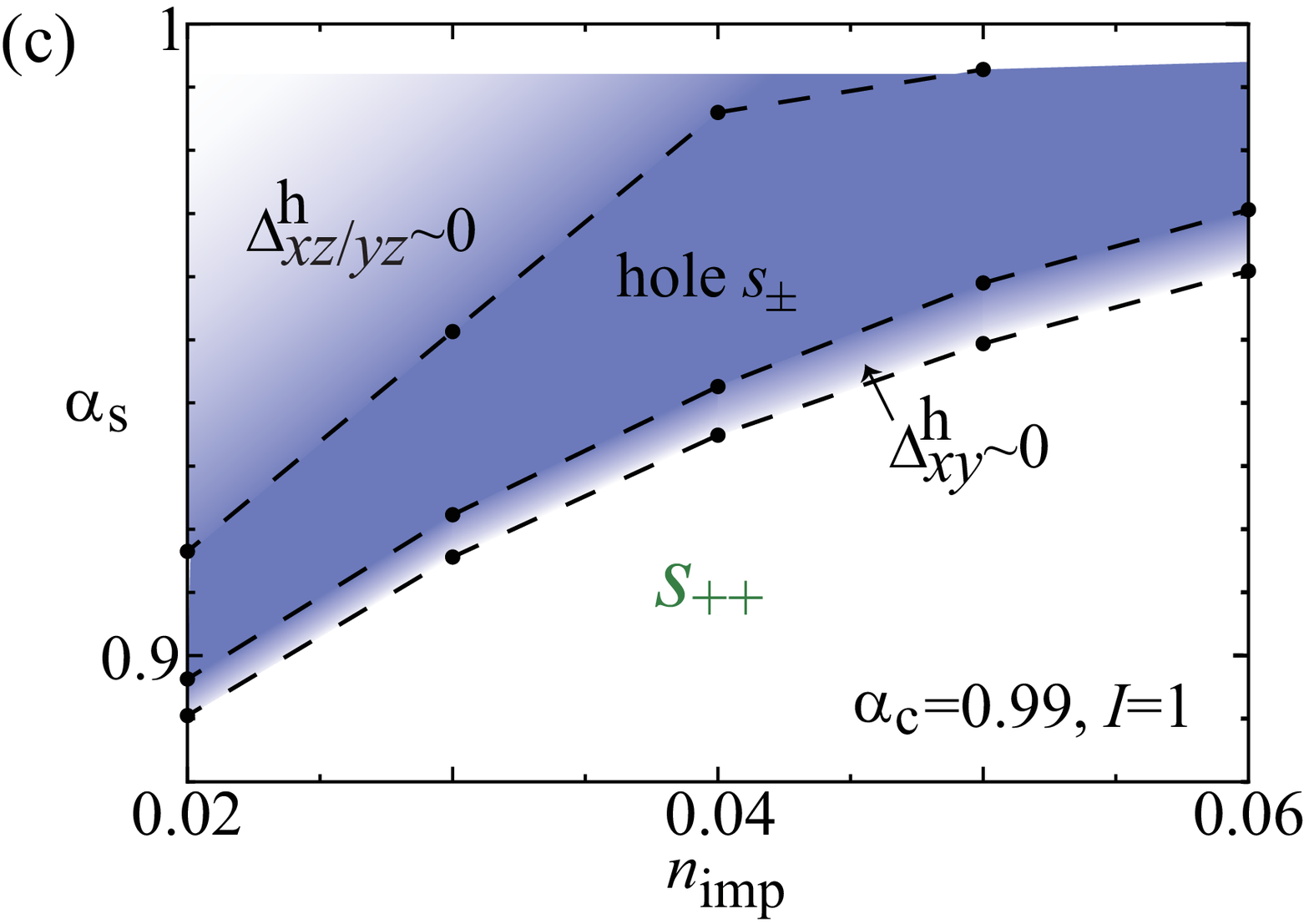}
\caption{
(Color online)
(a) The FSs of the three dimensional model for BaFe$_2$As$_2$
in the $k_z = 0$ plane.
The green, red, blue, and black lines correspond to 
$xz$, $yz$, $xy$, and $z^2$ orbitals, respectively.
(b) The obtained $\alpha_c$-$\alpha_s$ phase diagram.
The gap structure at each point {\bf a} $\sim$ {\bf c} is shown in the figure.
The horizontal axis is $\theta = \tan ( \bar{k}_y / \bar{k}_x)$,
where $( \bar{k}_x, \bar{k}_y )$ is the momentum on the FS with the origin at the center of each pocket.
(c) $n_{\rm imp}$-$\a_s$ phase diagram in the superconducting state.
The hole-$s_{\pm}$ state is realized for wide region in the phase diagram.
}
\label{fig:PB-Ba122}
\end{figure}

In this appendix,
we show that the hole-$s_{\pm}$-wave can be realized in Ba122 compounds
under the coexistence of orbital- and spin-fluctuations.
To construct the bandstructure,
we perform a local-density-approximation (LDA) band calculation for
BaFe$_2$As$_2$ using the WIEN2K code 
based on the experimental crystal structure.
Next, we derive the ten-orbital tight-binding model 
using the WANNIER90 code and the WIEN2WANNIER interface \cite{wannier90}.
However, the $d_{xy}$-orbital h-FS given by the LDA is much smaller
than the reports by ARPES studies.
In order to magnify its size,
we introduce the additional dependent potential $\Gamma$-Z line:
$H_{\mathrm{kin}} = H^0 + \sum_{l \sigma \bm{k}} e_{l}\frac12( \cos k_x \cos k_y + 1)
c^{\dagger}_{l \sigma} (\bm{k}) c_{l \sigma} (\bm{k})$,
where $e_l$ is the energy shift of the orbital $l$ in the $k_z$-axis.
We put $e_{xy} = 0.15$ eV, and the others are 0.
The obtained FSs in the $k_z = 0$ plane are shown in 
Fig. \ref{fig:PB-Ba122} (a).
Near the $\Gamma$ point, the $d_{xy}$-orbital h-FS is largest of all h-FSs.
Here, we neglect the SOI since the SOI-induced modification 
of the FSs is very small in Ba122 systems.

Next, we solve the Eliashberg equation
by introducing dilute concentrations of impurities
($I=1$ eV and $n_{\mathrm{imp}} = 0.04$) 
by applying the $T$-matrix approximation \cite{Saito-loop}
to make the obtained gap structures smoother.
Figure \ref{fig:PB-Ba122} (b) shows the obtained $\alpha_c$-$\alpha_s$ 
diagram of the superconducting state.
The gap functions on the FSs (in the $k_z = 0$ plane)
at each point {\bf a} $\sim$ {\bf c} are shown in the figure.
The obtained phase diagram is qualitatively similar to that 
for LiFeAs shown in Fig. \ref{fig:PB}.
Note that the gap structure in the region 
''$\Delta_{xy}^{\mathrm{h}} \sim 0$ ($\Delta_{xz/yz}^{\mathrm{h}} \sim 0$)''
is similar to the $s_{++}$ ($s_{\pm}$)-wave state.

In the hole-$s_{\pm}$-wave state at point {\bf b}, 
only the gap function on the outer h-FS ($d_{xy}$-orbital) is sign-reversed.
We stress that the gap on the inner e-FS takes the minimum
at $\theta = \pi/2$ in the $k_z = 0$-plane,
which is consistent with ARPES measurement for BaFe$_2$(As,P)$_2$ 
\cite{Yoshida}.
This hole-$s_{\pm}$-wave state is changed to the $s_{++}$-wave state 
(point {\bf c}) by increasing the impurities.
During this change, large impurity-induced DOS appears
at the Fermi level due to the outer h-FS,
which may correspond to the experimental 
disorder-induced gap structure in BaFe$_2$(As,P)$_2$ 
reported in Ref. \cite{Mizukami}.

Figure \ref{fig:PB-Ba122} (c) shows the $n_{\rm imp}$-$\a_s$ 
phase diagram in the superconducting state,
where $n_{\rm imp}$ is concentration of 
the non-magnetic impurity ($I=1$eV).
The hole-$s_{\pm}$ state is realized for wide region in the phase diagram,
reflecting the fact that the hole-$s_{\pm}$ state is realized by 
the ``cooperation'' between orbital- and spin-fluctuations.
That is, the ``intra-orbital repulsive interaction by spin-fluctuations''
mainly works between outer h-FS and e-FSs, whereas
the ``inter-orbital attractive interaction by orbital-fluctuations''
mainly works between inner and middle h-FSs ($d_{xz/yz}$-orbitals) and e-FSs.
In the SC-VC$_\Sigma$ theory, both $s_{++}$ and hole-$s_{\pm}$ states 
are realized in the case of $\a_c\sim\a_s$ even for $n_{\rm imp}=0$,
since the electron-orbiton coupling constant ($\sim U'$)
is enlarged by the vertex corrections for the gap equation ($\Delta$-VC)
\cite{Onari-SCVCS}.
A crossover from hole-$s_{\pm}$ state to $s_{++}$ state
may be realized in BaFe$_2$(As,P)$_2$ by electron irradiation
in Ref. \cite{Mizukami}.

\section{Spin susceptibility in the presence of the SOI}

In the main text, we studied the superconducting state
in LiFeAs in the presence of the SOI.
However, we neglected the SOI in the spin and charge susceptibilities
in the pairing interactions ${\hat V}^s(\q)$ and ${\hat V}^s(\q)$, respectively,
given in Eq. (\ref{eqn:Worb-cs}).
This approximation is justified if the relation 
$\chi^s_x(\q)\approx \chi^s_y(\q)\approx \chi^c_z(\q)$ is satisfied,
although this relation is not exact in the presence of the SOI.
In order to verify this relation,
here we calculate the spin susceptibilities in the presence of the SOI
using the RPA.

In the presence of the SOI, the spin is not conserved.
Therefore, the Green function is given by the following 
$20\times20$ matrix form,
${\hat G}(\k)=(i\e_n+\mu-{\hat h}(k))^{-1}$,
where ${\hat h}(k)$ is given in Eq. (\ref{eqn:hk}).
Then, the irreducible susceptibility is 
${\hat \chi}^{0}_{l\s,l'\s';m\rho,m'\rho'}(\q)
= -T\sum_{n,\k} G_{l\s,m\rho}(q+k) G_{m'\rho',l'\s'}(k)$.

\begin{figure}[!htb]
\includegraphics[width=0.99\linewidth]{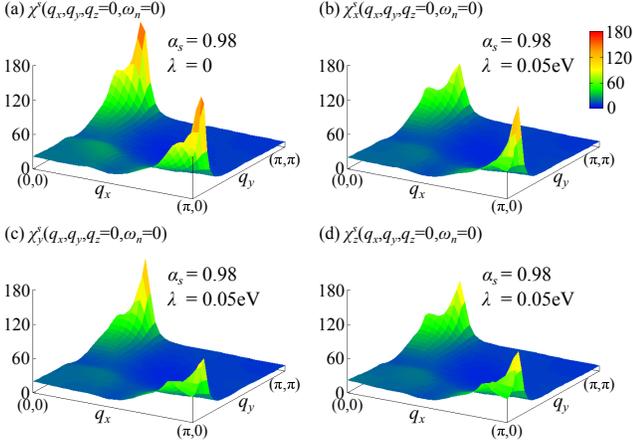}
\caption{
(Color online)
Spin susceptibilities in the case of $\a_S=0.98$.
(a) $\chi^s(\q)$ for in the absence of the SOI.
(b) $\chi^s_x(\bm q)$, (c) $\chi^s_y(\bm q)$ and (d) $\chi^s_z(\bm q)$
in the presence of the SOI ($\lambda=0.05$ eV).
}
\label{fig:chiS-SOI1}
\end{figure}

\begin{figure}[!htb]
\includegraphics[width=0.99\linewidth]{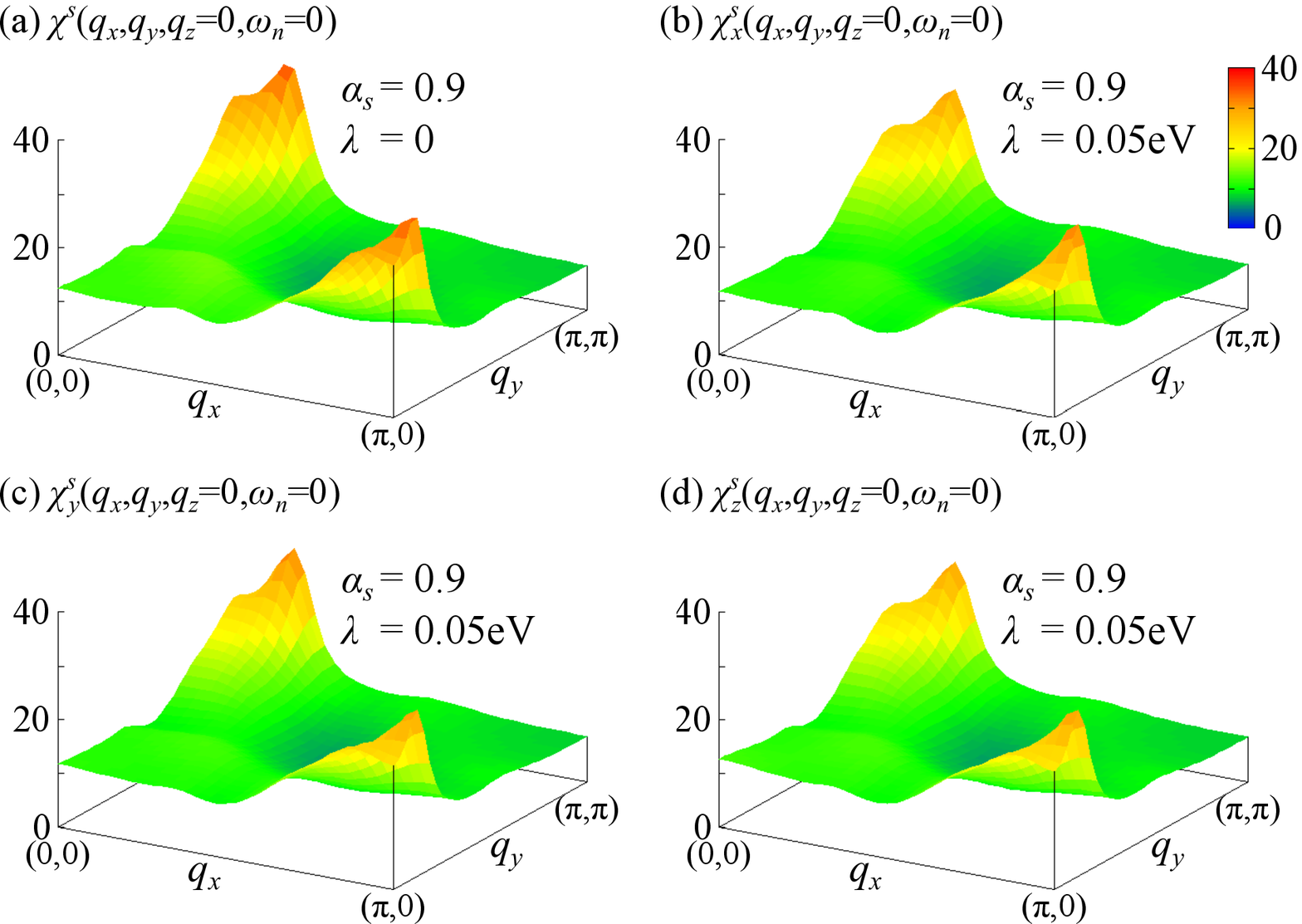}
\caption{
(Color online)
Spin susceptibilities in the case of $\a_S=0.90$.
(a) $\chi^s(\q)$ for in the absence of the SOI.
(b) $\chi^s_x(\bm q)$, (c) $\chi^s_y(\bm q)$ and (d) $\chi^s_z(\bm q)$
in the presence of the SOI ($\lambda=0.05$ eV).
}
\label{fig:chiS-SOI2}
\end{figure}

In the RPA, the susceptibility in the presence of the SOI is given 
by the following $20^2\times20^2$ matrix form:
\begin{eqnarray}
{\hat \chi}(\q)=(1-{\hat \chi}^0(\q){\hat \Gamma})^{-1}{\hat \chi}^0(\q),
\end{eqnarray}
where ${\hat \Gamma}$ is the Coulomb interaction in the matrix form:
$\Gamma_{l\s,l'\s';m\rho,m'\rho'}= \Gamma^c_{ll',mm'}\delta_{\s,\rho'}\delta_{\s',\rho}
+\Gamma^s_{ll',mm'}{\bm \s}_{\s,\rho'}\cdot {\bm \s}_{\s',\rho}$.
Then, the spin susceptibility in the $\mu$-direction is 
\begin{eqnarray}
\chi^s_\mu(\q)= \frac14 \sum_{l,m,\s,\s',\rho,\rho'}
\chi_{l\s,l\s';m\rho,m\rho'}(\q)\s^\mu_{\s',\s}\s^\mu_{\rho,\rho'}
\end{eqnarray}
where $\mu=x,y,z$.

In Fig. \ref{fig:chiS-SOI1}, we  show the numerical results 
in the vicinity of the magnetic critical point, at which 
the spin Stoner factor is $\a_S=0.98$.
Figure \ref{fig:chiS-SOI1} shows the obtained spin susceptibilities
for $\lambda=0$ ($U=0.439$ eV) in (a) 
and for $\lambda=0.05$ eV ($U=0.446$eV) in (b)-(d), respectively.
In the case of $\a_S=0.98$, the effect of the SOI on the 
spin susceptibility is not small.
In cotrast, we verified that the effect of the SOI on the charge
susceptibility 
$
\chi^c_{l,l';m,m'}(\q)= \sum_{\s,\rho}\chi_{l\s,l\s;m\rho,m\rho}(\q)
$
is very small even for $\alpha_C=0.98$.

In Fig. \ref{fig:chiS-SOI2}, we show the numerical results 
when the spin fluctuations are moderate ($\a_S=0.90$),
which is consistent with the weak spin fluctuations observed in LiFeAs.
Figure \ref{fig:chiS-SOI1} shows the obtained spin susceptibilities
for $\lambda=0$ ($U=0.403$ eV) in (a) 
and for $\lambda=0.05$ eV ($U=0.409$ eV) in (b)-(d), respectively.
In the case of $\a_S=0.90$, the relation 
$\chi^s_x(\q)\approx \chi^s_y(\q) \approx \chi^s_z(\q)$ 
is approximately satisfied, so the rotational symmetry of the 
spin susceptibility is almost satisfied.
Thus, the effect of the SOI on the pairing interaction 
is expected to be negligible.

Experimentally, LiFeAs is not close to the magnetic critical point 
because the observed spin fluctuations are small, 
so the relation $\a_S \sim0.9$ is expected in LiFeAs. 
We have verified that the gap structure in the $s_\pm$-state
shown in Fig. \ref{fig:gap2} is essentially unchanged even for $\a_S=0.90$,
similarly to our previous study in Ref. \cite{Saito-LiFeAs}.
Therefore, the effect of the SOI on the pairing interaction 
${\hat V}^{s,c}(\q)$ in LiFeAs is negligible, 
so the obtained results in the main text are essentially 
unchanged even if the SOI is included in the pairing interaction.


\end{document}